# Molecular interaction volume model of mixing enthalpy for molten salt system: An integrated calorimetry-model case study of LaCl$_3$-(LiCl-KCl)

Vitaliy G. Goncharov[1,2], William Smith[1], Jiahong Li[1], Jeffrey A. Eakin[1], Erik D. Reinhart[1], James Boncella[1], Luke D. Gibson,[3] Vyacheslav S. Bryantsev,[4] Rushi Gong[5], Shun-Li Shang[5], Zi-Kui Liu[5], Hongwu Xu[6,7], Aurora Clark,[1,8,*] Xiaofeng Guo[1,2,*]

[1] *Department of Chemistry, Washington State University, Pullman, Washington, 99164, United States*

[2] *Materials Science and Engineering Program, Washington State University, Pullman, Washington 99164, United States*

[3] *Computational Sciences and Engineering Division, Oak Ridge National Laboratory, P.O. Box 2008, Oak Ridge, TN 37831, United States*

[4] *Chemical Science Division, Oak Ridge National Laboratory, P.O. Box 2008, Oak Ridge, TN 37831, United States*

[5] *Department of Materials Science and Engineering, Pennsylvania State University, University Park, PA 16802, United States*

[6] *Earth and Environmental Sciences Division, Los Alamos National Laboratory, Los Alamos, New Mexico, NM 87545, United States*

[7] *Arizona State University, Tempe, AZ, United States*

[8] *Department of Chemistry, The University of Utah, Salt Lake City, Utah 84112, United States*

[*]E-mails: aurora.clark@utah.edu, x.guo@wsu.edu

**Abstract:** Calorimetric determination of enthalpies of mixing ($\Delta H_{mix}$) of multicomponent molten salts often employs empirical models that lack parameters with clear physical interpretation (e.g., coordination numbers, molar volumes, and pair potentials). Although such physics informed models are not always needed, a thermodynamic understanding of the relationships between excess energies of mixing and local to intermediate solvation structures is particularly important for pyrochemical separation, as is the case for lanthanides (Ln), which are common neutron poisons and critical industrial elements found in spent nuclear fuels. Here we implement the molecular interaction volume model (MIVM) to synthesize information from experimentally measured $\Delta H_{mix}$ (using high temperature melt drop calorimetry) and the distribution of solvation structures from *ab initio* molecular dynamics (AIMD) simulations. This was demonstrated by a case study of molten salt system consisted of $LaCl_3$ mixing with a eutectic LiCl-KCl (58mol% – 42mol%) at 873 K and 1133 K. The parameters modelled from MIVM were used to extrapolate excess Gibbs energy ($\Delta G_{mix}$), and compositional dependence of $La^{3+}$ activity in the $LaCl_3$-(LiCl-KCl) system. In contrast, by AIMD or polarizable ion model (PIM) simulations, a significant deviation regarding the predicted $\Delta H_{mix}$ was seen if computed directly from the molecular dynamic trajectories. The integrated experimental and simulation data within the MIVM formalism are generalizable to a wide variety of molten salts and demonstrate a significant improvement over currently employed methods to study molten salts for nuclear and separations sciences.

# 1 Introduction

Molten salt pyrochemical processing is an effective separation technique for actinides (An) and lanthanides (Ln), which is often a challenge due to their chemical similarity.[1–4] Its implementation and optimization require fundamental thermodynamic understanding of salt properties. Of particular interest is the excess Gibbs energy, which derives from the solvation structure of the mixed components. The dominating term enthalpy of mixing ($\Delta H_{mix}$) can be experimentally accessed by high temperature drop calorimetry,[5–12] and is also effectively studied by first-principle calculations, such as *ab-initio* molecular dynamics (AIMD) based on density functional theory (DFT), providing information on metal chemistry and speciation (e.g., coordination number and interatomic pair potential).[13–18] Although various thermodynamic models are available to couple thermochemical data (e.g., $\Delta H_{mix}$) and phase equilibrium data based on the Calculation of Phase Diagrams (CALPHAD) method,[19,20] most of these models include fitting parameters to be optimized, rather than being directly obtained from experiments or simulations. Given the potential of $\Delta H_{mix}$ to link solvation structure from the atomic scale to thermochemical properties at the macroscopic level, a model with chemically intuitive parameters would have unique advantages for developing the fundamental understanding needed for rational design of salt composition and operational conditions to tailor separation. In this work, we showcase an integration of the physically interpretable model (molecular interaction volume model, MIVM [21,22]), with experimental calorimetry and computational efforts (AIMD and the polarizable ion model (PIM) simulations[23–25]), and its application to reveal the structure-energetic landscape of a molten chloride salt system.

The pseudo-binary LaCl$_3$-(LiCl-KCl) molten salt has been studied, where (LiCl-KCl) is in its eutectic composition (~ 58 mol% LiCl and ~42 mol% KCl) with a melting temperature of 773 K.[26] When molten, eutectic LiCl-KCl can provide anodic dissolution of metallic spent nuclear fuel (SNF) during pyroprocessing, with An and Ln of interest recovered on the cathode surface via electrochemical deposition through applications of appropriate reductive potentials.[27] Molten salt pyrochemical processing offers a promising solution to separate metallic SNF from an open fuel cycle (used by majority of countries with active nuclear energy programs) [28,29], and to cycle up to 96% of critical metals, such as U back to fresh fuels[30–32], as well as a number of fission products that have industrial significance. In this work, LaCl$_3$ was selected as a constituent salt to be mixed with eutectic LiCl-KCl because they constitute ~1/3 of the fission products in SNF[1] and they present as a major challenge.[27,33,34] Uncertainty of the speciation and associated stability within the melt has deleterious effects on the efficiency and effectiveness of An-Ln seperation.[35] On the other hand, extraction of Ln from SNF provides an alternative source for critical metals rather than from rare-earth element (REE) deposits ore which often only enrich light REE and are commonly accompanied with high Th or U contents.[36] With high demands of separating Ln from SNF,[37–40] it is key to understand speciation and thermodynamics of La (as a surrogate for Ln in general) in eutectic LiCl-KCl.

Although prior calorimetric investigations of $\Delta H_{mix}$ of molten Ln chlorides with alkali/alkaline-earth metal chlorides have been reported,[41–47] the origins of the thermodynamic non-ideality (which may be due to the formation of complexes, oligomers, or metal-chloride networks induced by the "guest" metal salt component) are not well understood.[48,49] The correlation between solvation structure and energetics are not straightforward and rely heavily upon parameterized empirical models; the most commonly used are the associated solution model (ASM)[41–47] and the surrounded ion model (SIM)[4]. ASM assumes that the irregularity of mixing behavior may be expressed as a linear combination of regular mixing interactions ($\Delta_{mix}H^{REG}$) and locally ordered "associates" ($\Delta_{mix}H^{ASSOC}$).[50] From ASM, one can further derive the interaction parameter ($\lambda$) by $\lambda = \Delta H_{mix}/x \cdot (1-x)$ with $x$ being the primary mixed chloride (e.g., LnCl$_3$).[50] $\lambda$ can further be deconvoluted into a linear combination of coulombic and polarization interactions and the formation enthalpy of the associates. In assuming small deviations from regular solution behavior and ignoring solvation beyond first cation-anion coordination, ASM is quantitatively inadequate to describe some spectroscopic and computational results.[51] Alternatively, the SIM model is largely configuration based and primarily considers the charge of the mixing ions, with charge unsymmetric cations substitution yielding vacancies on the corresponding sublattices.[52] It allows for fitting of multicomponent salt systems, although the number of charge asymmetries is limited (i.e., trivalent + monovalent salts). SIM also does not reflect directly solvation structure,[51,53] and is subject to overfitting. These models demonstrate that there is significant opportunity to develop models with quantitative correlation of $\Delta H_{mix}$ to solvation structure and chemistry. On the other side with the computational efforts, a completely different approach entails the computation of $\Delta H_{mix}$ directly from the molecular dynamics (MD) trajectory obtained using either an empirical model (such as PIM) or DFT, typically employed at the generalized gradient approximation (GGA) for the exchange-correlation energy. While this approach does not introduce any additional empirical parameters beyond those used to describe the underlying computational method, it was noted that such 'direct methods' lack sufficient accuracy and often overpredict the magnitude of $\Delta H_{mix}$ for molten salts systems containing multivalent metal ions (e.g., Be$^{2+}$, U$^{3+}$, U$^{4+}$, Th$^{4+}$, etc.)[54–56].

Toward this end, we studied the mixing of LaCl$_3$ with 58mol% LiCl – 42mol% KCl eutectic melt using experimental calorimetry, together with AIMD and PIM modeling of $\Delta H_{mix}$ using the 'direct method', followed by implementing a modified MIVM for integrated data analysis. Within the calorimetry, a measurement temperature of 873 K for the low La content salts was chosen due to similar condition used in pyroprocessing [27], while a higher temperature of 1133 K was used to melt and access the high La content salts and to perform AIMD and PIM simulations across the whole LaCl$_3$-(LiCl-KCl) composition range.

An additional set of AIMD simulations was conducted at 1200 K, from which we obtained the radius distribution functions (RDF) and potential of mean force (PMF) of various metals, as well as the distribution of metal coordination number (CN) as a function of La content. The combined results, including enthalpies of mixing and solvation structures, were integrated to a modified MIVM mixing model that uses physically-based parameters that are either experimentally measurable and/or computationally available. Previously, MIVM has been used to explore the thermodynamics of mixing of binary, ternary, and high-order alloys.[57,58] With modified MIVM in this work, we also demonstrated the use of MIVM for the first time on the molten salt system. The parameters used by the MIVM are all physical and include the coordination number, pair potential, and molar volume, all of which can be readily determined by computation or experiment,[21,22,57] and thus will be very useful to be deployed more widely for thermodynamic studies in the molten salt community.

## 2 Methods

### 2.1 MIVM and its extension to model mixing enthalpy of molten salts

MIVM was originally formulated by Tao[22] to describe the excess thermodynamic properties of multi-component liquid systems based on non-random migrating local compositions (molecular cells) formed by the interactions of mixed components. The original model (described in SI section 1 and **Figure S1**) has not been implemented in a multicomponent molten salt system. We applied a pseudo binary formulation of MIVM with a modified definition of coordination number to the mixing of LaCl$_3$ in eutectic LiCl-KCl, described by equation 1; here the component 1 is LaCl$_3$ and the component 2 is eutectic LiCl-KCl:

$$\frac{\Delta H_m^M}{RT} = \frac{1}{2}\left\{\sum_{i=1}^{2} Z_i x_i \times \left[\left(\frac{\sum_{j=1}^{2} x_j B_{ji} \ln B_{ji}}{\sum_{j=1}^{C} x_j B_{ji}}\right)^2 - \left(\frac{\sum_{j=1}^{2}(1 + \ln B_{ji})x_j B_{ji} \ln B_{ji}}{\sum_{j=1}^{C} x_j B_{ji}}\right)\right]\right\}$$
$$- \sum_{i=1}^{2} x_i \left(\frac{\sum_{j=1}^{2} x_j V_{mj} B_{ji} \ln B_{ji}}{\sum_{j=1}^{2} x_j V_{mj} B_{ji}}\right) \quad (1)$$

where $R$ is the gas constant (kJ/mol·K), $T$ is the temperature (K), $Z_i$ is the liquidus first shell coordination number of component $i$, $V_{mj}$ is the liquidus molar volume of component $j$ (cm$^3$/mol), $B_{ij}$ and $B_{ji}$ are the pair potential parameters of the $i-j$ pairs defined as

$$B_{ij} = e^{-(\varepsilon_{ij} - \varepsilon_{jj})/RT}, \quad B_{ji} = e^{-(\varepsilon_{ji} - \varepsilon_{ii})/RT} \quad (2)$$

where $\varepsilon_{ii}$, $\varepsilon_{jj}$, and $\varepsilon_{ij}$ (kJ/mol) refer to the potential energies of the $i-i$, $j-j$, and $i-j$ pairs, respectively.[22]

In *equation 1*, $Z_1$ corresponds to the first cation shell CN of $La^{3+}$ (La – La) in molten $LaCl_3$, $Z_2$ corresponds to the averaged first cation shell CN of $Li^+$:$K^+$ in eutectic LiCl–KCl (0.58Li-K + 0.42K-Li). $V_{m1}$ and $V_{m2}$ correspond to the liquidus molar volumes of $LaCl_3$ and eutectic LiCl–KCl, respectively.[21] The potential energies[22] of the 1 – 1, 2 – 2, and 1 – 2 components pairs are likewise defined to correspond to the pair potentials of La–Cl–La ($\varepsilon_{11}$), Li:K–Cl–Li:K ($\varepsilon_{22}$), and La–Cl–Li:K ($\varepsilon_{12}$), respectively. Pair potential parameters, $B_{12}$ and $B_{21}$, can then be obtained using *equation 2*.[21]

## 2.1 Calorimetry methods

### 2.1.1 Differential Drop Calorimetric (DDC) Method

An argon sealed ampule of ultra-dry $LaCl_3$ was transferred to and opened within a glovebox ($O_2 < 1$ ppm, $H_2O < 0.5$ ppm). $LaCl_3$ beads were then subsequently ground into a fine powder and pressed into ~20 mg pellets using a hand die. The pellets were then loaded within into an annealed 3D printed airtight dropper,[59] which was used to transfer the sample to the calorimeter. All the samples' preparations, loading and storage was done within the glovebox. The $LaCl_3$ pellets were then dropped from room temperature into the calorimetric chamber (ambient exposure of < 1 s) which contained a nickel crucible with the molten LiCl-KCl eutectic (~200 mg) at 873 K in a steady argon environment. Upon dissolution and mixing of $LaCl_3$ within the LiCl-KCl eutectic, continuous drop enthalpy ($\Delta H_{cd}$) was obtained. The resulting $\Delta H_{cd}$ values were then deconvoluted using thermochemical cycles in **Table S1** yielding differential (or incremental) enthalpy of mixing ($dH_{i,mix}$) which denotes the changes in the heat of mixing corresponding to changes in La concentrations within the melt. $LaCl_3$ samples were dropped into the LiCl-KCl eutectic at periodic intervals (~1.25 hrs.) changing the mol% of La within the melt, thus generating $dH_{i,mix}$ values across the liquidus range of the $LaCl_3$ – LiCl-KCl system at 873 K. The $dH_{i,mix}$ values were then cumulatively summed to obtain molar enthalpies of mixing ($\Delta H_{mix}$) (**Table S1**), corresponding to the overall La concentration introduced into the eutectic LiCl-KCl melt. Validity of this method was evaluated based on the consistency of the $\Delta H_{mix}$ obtained from four independent trials reported in **Table S2**, with a curve presented in **Figure 1**.

### 2.1.2 Integral Drop Calorimetric (IDC) Method

$LaCl_3$ and LiCl-KCl powders were mixed in specific $LaCl_3$ – LiCl:KCl ratios and homogenized using an agate pestle and mortar yielding 5 physical mixtures with $LaCl_3$ concentrations corresponding to 1.77, 5.09, 7.06, and 19.26 mol% $LaCl_3$ measured at 873 K. Additionally, 4 additional mixtures were prepared with LaCl$_3$ concentrations corresponding to 28.68, 43.64, 68.95, and 82.36 mol% LaCl$_3$ and measured at 1133 K. Samples from the corresponding physical mixtures were then pressed into ~10 mg pellets using a hand dye and loaded into annealed 3D printed airtight droppers.[60] This approach is similar to literature work.[61,62] The physical mixtures were then dropped from room temperature into the calorimetric chamber which contained empty nickel crucibles at either 873 K or 1133 K in a stable argon environment (~5 ml/min flow). Upon introduction of the physical mixtures into the calorimetric chamber the pellets underwent the following thermal reactions within the crucibles: *i*) melting of the eutectic; *ii*) dissolution of LaCl$_3$ within the eutectic; and *iii*) mixing of the LaCl$_3$ – LiCl-KCl systems. Upon completion of these reactions, the integral heat effect, termed the drop enthalpy of physical mixtures ($\Delta H_{pd}$), was obtained. The resulting $\Delta H_{pd}$ values were then used in a corresponding thermochemical cycle (**Tables S3 and S4** for reactions at 873 and 1133 K, respectively) to calculate the molar $\Delta H_{mix}$ at a given physical mixture La content (**Table S5**). At least 3 consecutive measurements were made for a given physical mixture ratio, resulting in $\Delta H_{mix}$ values with experimentally bound uncertainties (**Table S5**). Data for the 873 and 1133 K trails are plotted as dark grey solid squares and solid diamonds in **Figure 1**, respectively, with two standard deviations serving as uncertainty bounds.

### 2.2 Computational methods

The CP2K package[63,64] version 9.1 using the Quickstep module was employed for Born-Oppenheimer density functional theory molecular dynamics simulations. The small box size of the simulations meant that small fluctuations in the cell dimensions result in large density changes, therefore experimental densities were used for the construction of the simulations. The pure salt densities were determined using the linear temperature dependent density formula:

$$\rho = a - bT \tag{3}$$

where the *a* and *b* parameters for the LiCl, KCl, and LaCl$_3$ came from the molten salt database by Janz.[65] The ideal mixing model developed at Idaho National Laboratory (Eq. 4) was implemented to determine the density of the molten salt mixtures.[66] Where the density, $\rho$, is given by the summation of the individual weight fractions and densities, $w_i$ and $\rho_i$, respectively.

$$\frac{1}{\rho_{salt,calc}} = \sum \frac{w_i}{\rho_i} \tag{4}$$

The simulation boxes were generated using PACKMOL,[67] the dimensions for the boxes ranged from (14.757 × 14.757 × 14.757 Å) to (20.45 × 20.45 × 20.45 Å), detailed information for the composition of each simulation can be found in the SI. The equilibrium procedure was executed in the *NVT* ensemble using a Nosé thermostat at 1200 K, higher than the reported melting point for this range of La concentration in a LiCl-KCl eutectic[68]. Geodecker, Teter, and Hutter (GTH) psuedopotentials with the Perdew, Burke, and Ernzerhof (PBE) exchange correlation function with D3 dispersion corrections were applied to all ions with accompanying DZVP-MOLOPT basis set for all ions.[69-73] It is well documented that the generalized gradient approximation functions in DFT poorly describe systems with strongly correlated *d*- and *f*-elements. This was resolved through the use of a Hubbard-like term implemented via DFT+$U$ with a $U_{eff}$ value of 4 eV.[74] All simulations were run in the NVT ensemble with a time step of 1 fs for a total of 150 ps with the first third discarded prior to analysis. The potential of mean force (PMF) was determined by the radial distribution function using $PMF(x) = -k_B T ln[P(x)]$, where the probability distribution $P(x)$ comes from the calculated RDF (shown in SI section 3, **Tables S7 and S8**, **Figures S2~S5**). Furthermore, additional computational detail for setting up PIM and AIMD simulations to directly compute the $\Delta H_{mix}$ are given in the SI section 4.

## 3 Results and Discussion

### 3.1 Computations of mixing enthalpy of LaCl₃ in eutectic LiCl – KCl from PIM and AIMD

The mixing enthalpies of molten salts can be computed directly from MD trajectories generated using the isothermal–isobaric ensemble (NPT) by averaging the normalized (per one molecular unit) potential energies across the composition range after subtracting the normalized potential energies of the end members, as follows:

$$\Delta H_{mix} = U_{AB} + (PV)_{AB} - x_A[U_A + (PV)_A] - x_B[U_B + (PV)_B], \tag{5}$$

where $U$, $P$, and $V$ denote the internal energy, the system pressure, and system volume, respectively. Subscripts $A$ and $B$ correspond to the two pure component systems (i.e., eutectic LiCl-KCl and LaCl₃), while $AB$ denotes mixtures with associated mole fractions, $x_A$ and $x_B$. At conditions close to the atmospheric pressure, the PV term is small and can be omitted. For example, in our PIM NPT (P = 1 bar) simulations, the PV contribution to $\Delta H_{mix}$ is less than 0.02 kJ/mol. Consistent with the previous studies of molten salts containing multivalent cations,[54-56] the PIM model reproduces the overall shape of $\Delta H_{mix}$ as a function of composition, but significantly overestimates the magnitude of $\Delta H_{mix}$, by almost 100% at the most negative $\Delta H_{mix}$ value (**Figure 1**). We note that the PIM model[23] for LaCl₃ was fitted to reproduce experimental liquid density and structure functions from the scattering experiments and was not specifically trained to yield accurate thermodynamic parameters for the interaction of the individual salt's components with its environment. The absolute error in $\Delta H_{mix}$ produced by the PIM model is < 6 kJ/mol, which is in the range of accuracy expected from such a model. This emphasizes the grand challenge of the computational methods to achieve 'chemical accuracy' of ~4 kJ/mol or better in predicting reaction thermochemistry and, as such, provide a more accurate description of chemical potentials and phase diagrams.

The AIMD simulations using the PBE-D3 density functional for the same compositions were attempt to directly reproduce $\Delta H_{mix}$. In this case, however, simulations were carried out in the canonical (NVT) ensemble, which are easier to converge than those in the NPT ensemble given limited AIMD sampling. A series of NVT simulations for different volumes were conducted to estimate the equilibrium density for each composition at the pressure near 1 bar (see **Figure S7**). In comparison to the experimental densities observed for the pure components, the PBE-D3 density functional underestimates the density of liquid LaCl$_3$ at T = 1133 K by 9.2%, yet provides an accurate density estimation for the LiCl-KCl eutectic mixture. The PIM and PBE-D3 densities deviate more as the mole fraction of LaCl$_3$ increases, with the PIM model accurately reproducing the density of liquid LaCl$_3$. In this respect, the deficiency of DFT is likely related to underestimation of the interaction of LaCl$_3$ with its environment, which is consistent with previous observations[51] that DFT-based AIMD simulations yielded smaller La-Cl CN and reduced first RDFs peaks for the La-Cl and La-La pairs compared to the PIM model. Mixing enthalpies using the AIMD NVT trajectories were computed or the cells with the smallest pressure magnitudes. The uncertainty in the $\Delta H_{mix}$ due to the average pressure deviating from 1 bar was estimated to be below 0.6 kJ/mol (**Figure S8**). In line with the PIM simulations, the magnitude of $\Delta H_{mix}$ from the AIMD simulations is significantly overestimated (more than twice) compared to the experimental values (**Figure 1**). In general, GGA DFT-based AIMD simulations are not expected to achieve a desirable level of 'chemical accuracy' for a diverse set of systems, especially for those containing f-block elements, where the self-interaction error of DFT becomes very prominent.[75] It remains to be seen if hybrid DFT methods can significantly reduce the error in $\Delta H_{mix}$, or resorting to more accurate methods, such as Random Phase Approximation (RPA), coupled cluster, and/or quantum Monte Carlo would be necessary to bring the error down to the level expected from CALPHAD modeling. Another promising option is the use of machine learning (ML) force field developments for testing these more advanced quantum mechanical methods for predicting thermodynamic properties of molten salts.

### 3.2 Application of MIVM to the mixing enthalpy of LaCl$_3$ in eutectic LiCl – KCl

**Table 1** provides the summary of input parameters for MIVM needed to construct $\Delta H_{mix}$ functions (**Figure 1**), in comparison with calorimetry-measured results. $V_{m1}$ and $V_{m2}$ were referenced to previous thermomechanical analysis (TMA) measurements.[76] Other four parameters ($Z_1$, $Z_2$, $B_{12}$, and $B_{21}$) were provided by AIMD or MIVM fitting. Specifically, the RDF results of La-La, Li-K, La-Li, and La-K (**Figure 2**) were used to calculate $Z_1$ and $Z_2$ (8.76 and 8.39, respectively). In our first attempt of applying MIVM, the pair potential energies $\varepsilon_{11}$, $\varepsilon_{22}$, and $\varepsilon_{12}$ were determined at the maximum positions of the cation-cation RDF curves (Index 1, **Figure 2**), which are also the global minima of the PMF. Using equation 2, $B_{12} = 1.11$ and $B_{21} = 1.09$ were derived and shown in **Table 2**. However, the PMF determined from this method may only include the effects of columbic interactions from separate ions in the melt which is an oversimplified approximation. The yielded "AIMD+MIVM" derived $\Delta H_{\mathrm{mix}}$ curve (**Figure 1**) deviates more endothermically than the experimental values (also see section 2 of the SI, **Tables S2 and S5**). Recognizing such deviation originated from PMF may also attributes to the finite size effect or a sensitivity to the distribution of species in the solution sampled by AIMD, the primary reason may come from the exclusion of other effects in cation-cation PMF that are representative of non-ideal local features possibly due to the formation of complexation or oligomerization.

To take into account of the non-ideal mixing effects, we have done two kinds of modification to better connect the solvation structure to energetics. First, the PMF-related $B_{ij}$ parameters were relaxed (with CN and molar volumes fixed) through linear regression of MIVM against experimental $\Delta H_{\mathrm{mix}}$ values (**Figure 1**), and determined to be $B_{12} = 1.48 \pm 0.02$ and $B_{21} = 0.98 \pm 0.02$ (**Table 1**). PMF corresponding to the new $B_{ij}$ from this "Calorimetry+MIVM" approach have values higher than those from "AIMD+MIVM", confirming that the averaged interatomic potentials represent a solvation structure that is beyond the first cation-cation shell environment. For instance, deconvolution of the Li-K and K-Li RDFs (**Figure 2b** and **Figure S2**, respectively) reveals that the Li − K of eutectic LiCl-KCl have two coordination environments. This has likewise been observed at similar temperatures by Emerson et al.[51] for the LaCl$_3$–NaCl molten chloride salt system, where La$^{3+}$ was found to occupy several shallow free energy minima, with little to no barriers separating these metastable states from the global minimum.

Thus, in the second modification approach ("AIMD+Calorimetry+MIVM"), we determined PMF at the average of two deconvoluted Li-K RDF peak positions (index 1' in **Figure 2b**) to be $\varepsilon_{22} = -3.92$ kJ/mol at 4.55 Å, in comparison to $\varepsilon_{22} = -5.85$ kJ/mol at 4.15 Å derived by the "AIMD + MIVM" approach. Similarly, the first La-Li cation-cation coordination can also be deconvoluted into two locally adjacent environments using normal distribution functions (**Figure 2c**), and yielded an averaged distance of 4.33 Å (index 1' in **Figure 2c**). PMF of La-Li at this position was determined to be $\varepsilon_{12} = -5.47$ kJ/mol, compared to -6.04 kJ/mol at the global PMF minima (4.13 Å). These two new PMF $\varepsilon_{22}$ and $\varepsilon_{12}$ led to new pair-potential parameters, $B_{12} = 1.38$ and $B_{21} = 1.04$, respectively. Because their derivations were inspired by the mismatch of measured $\Delta H_{\mathrm{mix}}$ and those reproduced by "AIMD+MIVM", we considered them to be values still obtained from AIMD but experimentally benchmarked. Again, applying new $B_{ij}$ in conjunction with fixed CN and molar volumes, we can more rationally generate $\Delta H_{mix}$ based on the solvation structure.

The "AIMD+Calorimetry+MIVM" model fully reproduced $\Delta H_{mix}$ within the experimental uncertainty (**Figure 1**). The slightly endothermic character implies the MIVM model only considers binary interactions, with non-trivial corrections needed for ternary or quaternary interactions. Similar type of results was observed in the prediction of partial Gibb's energy of the Cs – Sb – Pb alloys by Poizeau and Sadoway using the MIVM method.[77] However, unlike the aforementioned discussion, the more endothermic mixing energy predicted by MIVM was attributed to its inability to account for the first nearest neighbor interactions; whereas in our work on molten salt, missing high-order interactions is hypothesized to account for the discrepancy MIVM prediction, which will be elaborated more in the next section 3.4.

### 3.3 Sensitivity of MIVM to variances in input parameters

Knowing that MIVM can effectively describe and predict thermochemistry of pseudo-binary molten salt, we further examined the model sensitivity to variances of the input parameters. An iterative fitting procedure was performed through perturbations of $Z_1$, $Z_2$, $B_{12}$, and $B_{21}$ and compared to the experimental $\Delta H_{mix}$. This was accomplished by changing one of the parameters from **Table 1** (excluding $V_{m1}$ and $V_{m2}$) by ± 10% in magnitude while keeping all other parameters constant. The stability of model from these perturbations are presented in **Figure 3**. Perturbation in coordination number results in statistically insignificant impacts on MIVM. This also suggests that thermochemical model is relatively inert to variances in metal CN if any experimental errors present. On the other hand, it also suggests CN may not be reasonably constrained from MIVM. In contrast, variance in $B_{ij}$ resulted in much big differences (~25%) in generated $\Delta H_{mix}$ from ± 10% perturbations, suggesting a higher sensitive dependence. Accurate and rationale determination of the pair potential energies from the MIVM model is thus possible to be used with high confidence for benchmarking AIMD calculations.

Furthermore, to better distinguish the statistically significant difference between the curves produced in **Figure 3**, analysis of covariance (ANOVA)[78] was performed on the perturbated data sets (**Figure S5a**). ANOVA suggests that MIVM generated $\Delta H_{mix}$ is statistically different at the 0.05 level with data sets attained from ± 10% perturbation of $Z_2$, $B_{12}$, and $B_{21}$ parameters. Additionally, ANOVA revealed that when comparing the data sets resulting from related perturbated parameters (+ $Z_1$ vs. +$Z_2$, +$B_{12}$ vs. +$B_{21}$, -$Z_2$ vs. -$Z_1$ and -$B_{12}$ vs. -$B_{21}$) the distribution of the resulting values is not statistically different at the 0.05 level (**Figure 3c**). This suggests that if the paired input parameters have similar directional inaccuracies the resulting differences of MIVM fits will not be statistically distinguishable.

With the above model sensitivity analysis, we suggest that when experimental data are available the MIVM fit of measured $\Delta H_{mix}$ can be used (and perhaps the best experimental method) to constrain the pair potential energies. Reversely, the use of MIVM to extract coordination numbers should only be attempted when there is high confidence in the pair potential energies or when there is significant lack of experimental or computational inputs for the $Z_1$ and $Z_2$ parameters. This is exemplified by the calculation of impact on fit extracted from the interactive perturbation of each parameter (**Figure 3b**), which ranks $B_{ij}$ higher importance than $Z_i$ on perturbing $\Delta H_{mix}$ based on MIVM. This trend reiterates that the MIVM is robust to uncertainty in the coordination number of the first cation – cation shell.

### 3.4 Bridging $\Delta H_{mix}$ and the solvation structure of molten $LaCl_3$ – $LiCl$ – $KCl$

The MIVM fit of $\Delta H_{mix}$ produced an asymmetric mixing curve with the minimum of -5.45 kJ/mol at 41.71 mol% $LaCl_3$, in general agreement with previous studies of $LaCl_3$ mixing in alkali chlorides[4,79]. Particularly, the magnitude of the curve and the asymmetry closely resemble those of $LaCl_3$ in molten $NaCl$[4,79], likely due to the similarity of the ionic radii of $Na^+$ to the weighted mean of $0.58Li^+$ - $0.42K^+$ cations in the eutectic.[80] The irregularity of $\Delta H_{mix}$ can be explained by the AIMD simulated solvation structure of $LaCl_3$ in the molten eutectic $LiCl$-$KCl$. **Figure 4** shows the solvation environment of $La^{3+}$ at 1, 7, 20, and 100 mol% $LaCl_3$, and **Figure S5** shows the CN distributions of the first shell $La$–$Cl$ at corresponding $LaCl_3$ concentrations (**Table S8**). **Figures S3 and S4** present the evolution of RDF and CN of cation – cation and cation – anion pairs (respectively) at different $LaCl_3$ concentrations in the eutectic $LiCl$–$KCl$. With relevant CN and pair distances tabulated in **Tables S6 and S7** for cation-anion and cation-cation sets, respectively.

In the dilute $LaCl_3$ concentrations (i.e., $\leq$ 1 mol% $LaCl_3$) AIMD predicts that $La^{3+}$ forms dominantly monomeric locally ordered structures with $La$ primarily 6- and 7-fold coordination (67% and 32% in CN distribution, respectively) with the adjacent chlorides, forming $LaCl_6^{3-}$ and $LaCl_7^{4-}$ complexes. At these low concentrations, high $La$-$Cl$ CN (>7) is minimal with only 1% of $La$ coordinated by 8 $Cl$ in the first shell (**Figure S5**). $LaCl_6^{3-}$ has been previously hypothesized to form in alkali chloride melts,[50] with their formation further constituting the fundamental assumption of the ASM method. However, our AIMD results demonstrate that even at dilute $La$ loadings, there is an equilibrium of the $LaCl_6^{3-}$ and $LaCl_7^{4-}$ complexes. A recent work by Emerson et al. also demonstrated the coexisted complexes, considering both the past and recent spectroscopic and computational studies.[51]

As the concentration of $LaCl_3$ increasing to 7 mol%, more complex solvation environments for $La$ emerges, with the presence of $LaCl_x^{3-}$ monomers, dimers, and trimers with approximate populations of 85.98, 11.24, and 2.78%, respectively, in the equilibrium (**Table S8**). The distribution of $La$–$Cl$ CN of 6, 7, and 8 also changes to 55, 42, and 3%, respectively, with $LaCl_6^{3-}$ becoming less prevalent. This is likely due to the dimeric and trimeric formations within the melt which are able to stabilize La – Cl with higher coordination due to the ability of edge and corner sharing. Oligomerization continues at 20 mol%. The distribution of monomers, dimer, trimers, and oligomers ($x \geq 4$ in $La_x–Cl_z$) changes to 46.14, 14,75, 8.42 and 30.69%, respectively. At 20 mol%, the population of 7-coordinated La also increases to 45 %, comparable to that of 6-coordianted (46 %). Increasing CN and oligomerization are also previously observed in the related molten salt systems, such as $LaCl_3$-NaCl or $UCl_3/UCl_4$–NaCl.[51,53] Furthermore, RDF of La-Cl shows an increase of the average La–Cl CN to 6.58 at 7 mol% and 6.82 at 20 mol%, from 6.34 at 1 mol% (**Table S7**). The corresponding La–Cl distances elongate as well, by forming edge or corner sharing polyhedra to lower the polarizability induced on the shared Cl by the adjacent La, which together lead to the stabilization of the dimeric and trimeric La-Cl structures. As a result, these oligomers have higher La – Cl CN and lowered polarization of each coordinated Cl. Describing the energetic contribution of oligomers needs model beyond binary interaction. While in the MIVM method, although high-order interaction is not explicitly defined, its inclusion is a natural consequence of the "Calorimetry + MIVM" or the "AIMD + calorimetry + MIVM" approach. This is because the relaxed or benchmarked interatomic pair potential reflects an effective binary interaction that includes the mean-field effect of high-order interactions. The effective metal-metal distance (e.g., indices 1' in **Figure 2**), based on which the averaged interatomic pair potential is defined, should always be farther than the first metal-metal coordination peak, and represents the overall oligomer geometry.

The oligomeric networks continue expanding at high $LaCl_3$ loading. The interconnections of the La-La are likely driven by the deficiency of chlorides provided by LiCl-KCl in the system, requiring La to form complex chloride sharing networks.[13] At 100 mol% $LaCl_3$, AIMD suggests that its molten structure can be described by salt networks of interconnected La–Cl joined by edge, corner, and face sharing polyhedra. The La – Cl CN populations are 11, 48, 34, and 7% for the 6-, 7-, 8-, and 9-coordinated environments, respectively, with an average CN of 7.37 that is in general agreement with previous studies.

The above AIMD calculated solvation structures can thus rationalize thermochemistry measured by calorimetry, with the irregular mixing behavior of $\Delta H_{mix}$ attributing to the formation of complex and concentration-dependent La-metal coordinate environments. The exothermic enthalpies of mixing can then be further explained by relaxing of induced polarization to form various oligomer salt local structures. Previous calorimetric mixing studies on $LaCl_3$ in alkali chloride media demonstrate that as the polarization capacity of the spacer salt cation is increased, the resulting $\Delta H_{mix}$ becomes more exothermic.[20] This corresponds to the ability of larger alkali cations to better "share" the Cl and allow $LaCl_3$ to reach its most stable thermodynamic state across the entire melt composition.[81] Additionally, the larger alkali size of alkali metals such as K, Cs, and Rb allow for greater coordination shells which enables the softer cations to stabilize the La – Cl oligomeric structures in a charge alternating network.[81] Evidence of this can be observed by comparing the AIMD simulated La – Li versus La – K CN. Even though, there is more Li than K within our mixing system the RDF suggest that $CN_{La-Li} < CN_{La-K}$ at all examined $LaCl_3$ concentrations (**Table S7**). This may suggest that K preferentially interacts with the local chloride environment formed by the La, contributing to the non-random mixing behavior of $LaCl_3$ in LiCl-KCl. Besides polarization, the effects of Coulombic ordering and melt packing may likewise plays role in dictating the thermodynamically favored solvation state,[81] which requires further studies.

## 4 Conclusions

In this work, for the first time, we developed a thermodynamic benchmark by using calorimetrically measured enthalpy data for interatomic potential, which are difficult to be cross-checked by other experimental techniques. The commonly employed computational methods for modeling molten salts, such as PIM and PBE-D3, fall short in accurately predicting $\Delta H_{mix}$ directly from MD trajectories. The integration of calorimetry and first-principles calculations can be realistically achieved by MIVM, enabling an effective experiment-modeling loop for complete thermodynamic understanding of salt mixing. As a showcase study for the calorimetry-model integration, we performed calorimetric experiments to determine the enthalpy mixing functions of $LaCl_3$ in eutectic LiCl-KCl across the whole La concentration range at high temperatures. AIMD was performed to generate the corresponding solvation structures, and interatomic pair potentials. Using MIVM as the platform, we found that sole use of AIMD cannot predict thermochemistry and resulted in underestimation of the exothermic nature of mixing. Calorimetry-generated $\Delta H_{mix}$ can be used to fit PMF functions and thus effectively benchmark AIMD calculations with high sensitivity. This is because MIVM is robust to errors in coordination numbers but sensitive to errors of the pair potential parameters. Furthermore, the irregular mixing behavior of $\Delta H_{mix}$ attributing to the formation of complex and concentration-dependent La-metal coordinate environments that push the effective metal-metal distance, based on which the averaged interatomic pair potential is defined, forward from the first metal-metal coordination peak. The averaged interatomic pair potential thus represents the effective binary interaction, which also potentially includes high-order interactions that are not initially included in MIVM. This study suggests a novel approach to combine experimental thermodynamics with first-principles calculations for simultaneously revealing solvation structures and energies of molten salt systems.

## Acknowledgements

The authors acknowledge financial supports by the U.S. Department of Energy, Office of Nuclear Energy, Nuclear Energy University Programs via Awards No. DE-NE0009288 and DE-NE0009444. The early stage of the work was supported by Faculty Seed Grant to X.G. at Washington State University. Additional support was through collaboration, services, and infrastructure through the Nuclear Science Center User Facility at WSU, the WSU-PNNL Nuclear Science and Technology Institute, and Alexandra Navrotsky Institute for Experimental Thermodynamics. We are grateful to Rajni Chahal (ORNL) for her assistance in the setup of PIM simulations. The work at the Oak Ridge National Laboratory was supported by the Office of Materials and Chemical Technologies within the Office of Nuclear Energy, U.S. Department of Energy. This research used resources of the Oak Ridge Leadership Computing Facility, which is a DOE Office of Science User Facility, supported under Contract DE-AC05-00OR22725.

## Tables

**Table 1.** AIMD and previously determined experimental parameters used to estimate the $\Delta H_{mix}$ of $LaCl_3$ in LiCl-KCl eutectic through *equation 3* (curve 1 in **Figure 1**).

| Parameter | Magnitude |
|:---:|:---:|
| $Z_1^{\ddagger}$ | 8.76 |
| $Z_2^{\ddagger}$ | 8.39[*] |
| $V_{m1}^{\dagger\,76}$ | 70.27 $cm^3$/mol |
| $V_{m2}^{\dagger\,76}$ | 32.51 $cm^3$/mol |
| $B_{12}^{\ddagger}$ | 1.11 |
| $B_{21}^{\ddagger}$ | 1.09 |
| $B_{12}^{\ddagger\ddagger}$ | 1.48 |
| $B_{21}^{\ddagger\ddagger}$ | 0.98 |

[‡]Obtained from AIMD in this study. [*]Average value. [†]Experimentally obtained in previous studies. [‡]Values obtained through regression fitting of experimental $\Delta H_{mix}$ by the MIVM.

**Table 2.** AIMD determined distances between the cation pairs with the corresponding PMF energies. PMF energies are obtained at the energetic minimum of the cation-cation pairs.

| Paths | Distance (Å) | PMF (kJ/mol) |
|:---:|:---:|:---:|
| La – Cl – La[‡] | 5.05 | -5.98 |
| La – Cl – Li[†] | 4.13 | -6.04 |
| La – Cl – K[†] | 4.83 | -7.36 |
| Li – Cl – K[†] | 4.15 | -5.85 |
| Li – Cl – K[*] | 4.55 | -3.92 |

[‡] AIMD calculated using pure molten $LaCl_3$ at 1200 K, corresponding to $\varepsilon_{11}$. [‡]AIMD calculated using pure 58mol% LiCl – 42mol% KCl eutectic at 1200 K, corresponding to $\varepsilon_{22}$. [†]AIMD calculated using 20 mol% La loading in the $LaCl_3$- LiCl-KCl system 1200 K, corresponding to $\varepsilon_{12}$ components with the applied $\varepsilon_{12} = 0.58$(La-Cl-Li) + 0.42(La-Cl-K) [*]Experimentally benchmarked $\varepsilon_{22}$ pair potential.

## Figures

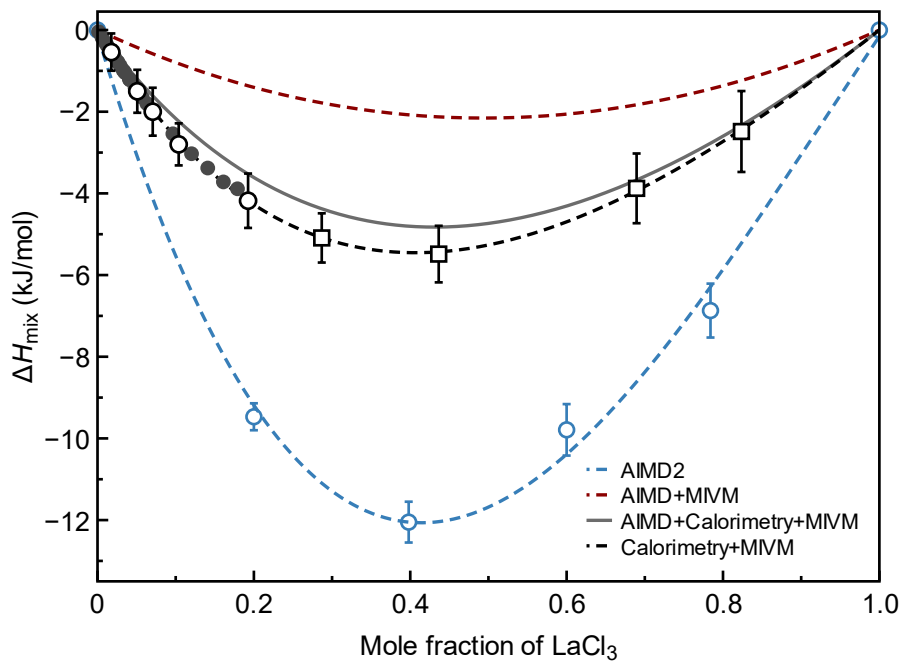

**Figure 1.** Experimentally and computationally obtained $\Delta H_{mix}$ in the present work. Closed and open circular data correspond to the experimental $\Delta H_{mix}$ values obtained through the continuous ($\Delta H_{cd}$) and physical mixture drop ($\Delta H_{pd}$) methods at 873 K; and the closed cubic data represent experimentally obtained $\Delta H_{mix}$ values through $\Delta H_{pd}$ at 1133 K. Dotted blue curve and data represent the direct computation of $\Delta H_{mix}$ from AIMD, thus labeled as AIMD2. Dotted red curve represents the $\Delta H_{mix}$ curve obtained based on inputting AIMD values to the MIVM model. Dotted black curve represents the MIVM regression fit of the experimental data with $B_{12}$ and $B_{21}$ as free parameters. Gray solid curve represents the $\Delta H_{mix}$ curve obtained by AIMD using MIVM and experimentally benchmarked $B_{12}$ and $B_{21}$ parameters.

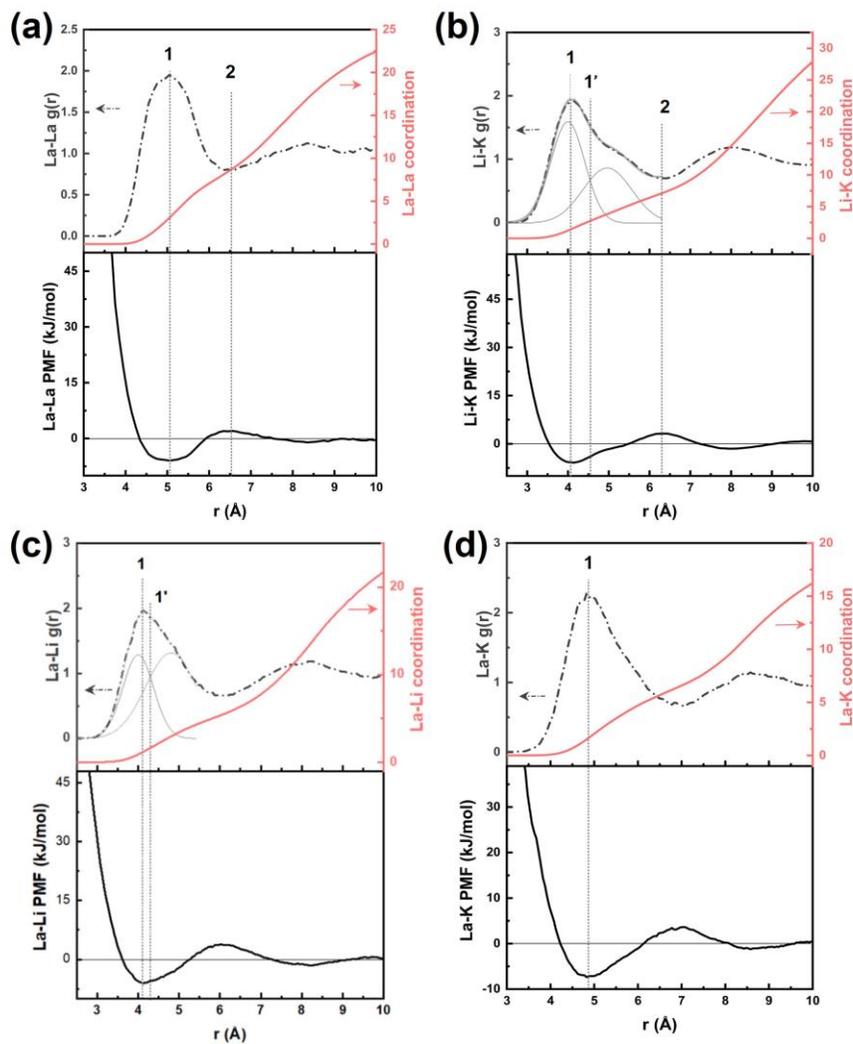

**Figure 2.** AIMD calculated radial distribution functions (RDF) depicted as the dashed grey curves, CN represented as solid red curves, and potentials of mean force (PMF) depicted as solid black curves, for cation-cation pairs in the $LaCl_3 – LiCl$-KCl system. (a) La-La pairs of pure molten $LaCl_3$. Dotted indexes marked by 1 and 2 represent the PMF energy for the $\varepsilon_{11}$ parameter, and $Z_1$ La-La CN for MIVM, respectively. (b) Li-K pairs of molten 58mol% LiCl - 42mol% KCl eutectic. Dotted indexes marked by 1 and 2 represent the PMF energy for the $\varepsilon_{22}$ parameter, and $Z_2$ Li – K CN for MIVM, respectively. Index marked by 1' corresponds to the PMF energy of the $\varepsilon_{22}$ and La-Li component of $\varepsilon_{12}$ parameter benchmarked by the MIVM fit of the experimental $\Delta H_{mix}$. K-Li RDF and CN are shown in **Figure S2**. (c) La-Li pairs of molten $LaCl_3 – LiCl$-KCl melt with 20mol% $LaCl_3$ loading, calculated at 873 K. Dotted index marked by 1 represents the La-Li PMF energy component for the $\varepsilon_{12}$ parameter for MIVM. (d) La-K pairs of molten $LaCl_3 – LiCl$-KCl melt at 20mol% $LaCl_3$ loading. Dotted index marked by 1 represents the La-K PMF energy component for the $\varepsilon_{12}$ parameter for MIVM.

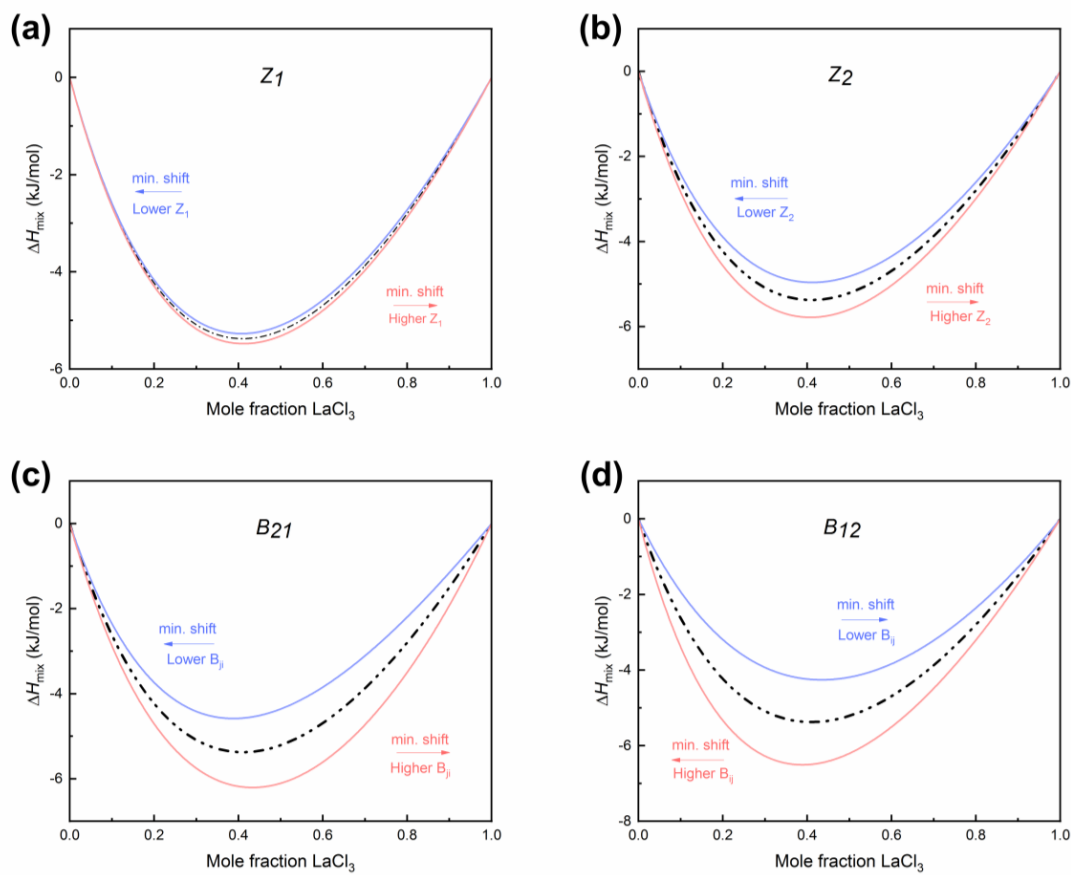

**Figure 3.** Iterative fitting analysis by MIVM fit of experimentally obtained $\Delta H_{mix}$ through single variable perturbation. Black dashed curve represents the MIVM fit of calorimetric data with values from **Table 1** ($Z_1^{\ddagger}$, $Z_2^{\ddagger}$, $B_{12}^{\ddagger}$ and $B_{21}^{\ddagger}$). The red solid curve represents raising the magnitude of the corresponding parameter by 10% and the blue solid curve represents lowering the magnitude of the corresponding parameter by 10%. (a) Perturbation of the $Z_1$ parameter; (b) Perturbation of the $Z_2$ parameter; (c) Perturbation of the $B_{21}$ parameter; and (c) Perturbation of the $B_{12}$ parameter.

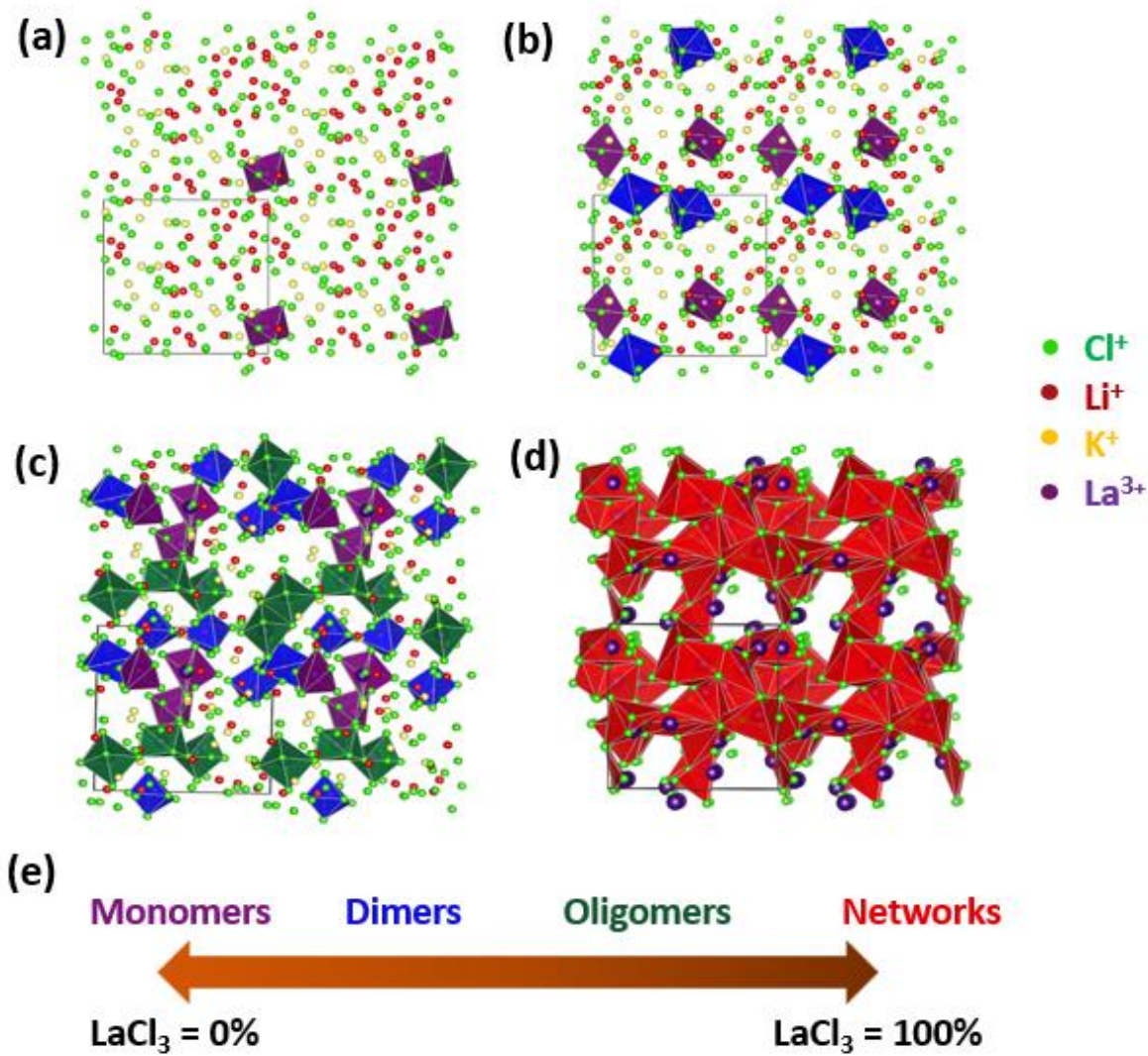

**Figure 4.** AIMD calculated solvation structures of LaCl₃ in 58mol% LiCl – 42mol% KCl eutectic, and the periodic boundary is outlined in black. Purple polyhedra represent the monomeric $LaCl_z$ species (ave. z = 6.34), blue polyhedra dimeric the $La_2Cl_z$ species, green polyhedra oligomer $La_xCl_z$ (x ≥ 3), and red polyhedra $La_xCl_z$ networks (x >>1000). La-Cl first shell coordination distributions are demonstrated in **Figure S5** and **Table S8**. (a) LaCl₃ speciation at 1mol% LaCl₃, calculated at 873 K;(b) LaCl₃ speciation at 7 mol% LaCl₃, calculated at 873 K;(c) LaCl₃ speciation at 20 mol% LaCl₃, calculated at 873 K;(d) Structure of pure LaCl₃ calculated at 1133 K; and (e) AIMD determined generalized trend of the LaCl₃ speciation in LiCl-KCl.